\documentclass[12pt,a4paper]{article}
\usepackage{epsfig}
\usepackage{a4}
\usepackage{graphics}
\usepackage{amsmath}
\usepackage{subfigure}

\newcommand{\half}{\mbox{${\textstyle \frac{1}{2}}$}}           
\newcommand{\threehalf}{\mbox{${\textstyle \frac{3}{2}}$}}      
\newcommand{\twothird}{\mbox{${\textstyle \frac{2}{3}}$}}       
\newcommand{\third}{\mbox{${\textstyle \frac{1}{3}}$}}          
\newcommand{\fourth}{\mbox{${\textstyle \frac{1}{4}}$}}         
\newcommand{\extrabrak}{\mbox{${\textstyle \frac{16}{9}}$}}     
\newcommand{\ninehalf}{\mbox{${\textstyle \frac{9}{2}}$}}       

\begin{document}
\setlength{\unitlength}{1mm}
\mbox{ }
%

\begin{flushright}
{\bf TSL/ISV-2005-0289 \\
February  2005}
\end{flushright}

\vspace{5mm}

\begin{center}

\vspace*{10mm}

{\Large{\bf CP violation in  $K^{\pm} \to \pi^0 \pi^0 \pi^{\pm}$
decay }}\\[8ex]
{\large G\"{o}ran F\"{a}ldt\footnote{Electronic address: faldt@tsl.uu.se}
\\[1ex]
{\normalsize  Division of Nuclear Physics,  Box 535, 751 21 Uppsala, Sweden}}
\\[2ex]
{\large  Evgeny Shabalin\footnote{Electronic address: shabalin@heron.itep.ru}
\\[1ex]
{\normalsize Institute for Theoretical and Experimental Physics, 117218 Moscow,
B.Cheremushkinskaya 25, Russia}}
\\[8ex]

\vspace{5mm}

{\bf Abstract}
\end{center}
CP violation leads to a difference between the parameters $g^+$ and
$g^-$ describing the energy distributions of the charged pions 
produced in the $K^+ \to
\pi^0 \pi^0 \pi^+$ and $K^- \to \pi^0\pi^0 \pi^-$ decays. We study the
difference $(g^+ - g^-)$ as a function of the relative 
contributions from the QCD-penguin  and the electroweak-penguin 
diagrams. We find that the
combination of these contributions in $(g^+ - g^-)$ is very similar
to the corresponding one defining the parameter 
$\varepsilon'$ in the $K_L \to 2\pi$ decays.
This observation allows a determination of the  value of $(g^+ - g^-)$, 
using  data on $\varepsilon'$.

\vspace{5mm}
\noindent

\noindent
PACS: 13.75 Jz, 14.20.-c, 25.10.+s
\vfill

\clearpage
\setcounter{page}{1}

\baselineskip 4ex

\newpage
\section{Introduction}
Since 1989 it has been known that the magnitude of direct CP
violation in the $K_L \to 2 \pi$ decays crucially depends 
on the relative strengths
of the imaginary parts of                        
the QCD-penguin  (QCDP) and the electroweak-penguin 
(EWP) contributions to the amplitude \cite{1}. 
The reason for this sensitivity
is that the contributions  to
$\varepsilon'$ from the two diagrams
have opposite signs and partially cancel one another.

 As the dynamical structures of the amplitudes for 
$K^{\pm} \to (3\pi)^{\pm}$ differ
from those for $K_L \to 2\pi$, there is no immediate relation
 between the
strengths of direct CP violation in the $K_L \to 2\pi$ and the
$K^{\pm}\to  (3\pi)^{\pm}$
decays. In particular, it was observed in refs.~\cite{2,3}
that in contrast to
the situation in $K_L \to 2\pi$, 
the CP violating effect
in $ K^{\pm} \to \pi^{\pm} \pi^{\pm} \pi^{\mp}$
produced by the QCDP contribution is enhanced by the EWP contribution.

However, in the present note, we shall demonstrate
 that  the $K^{\pm} \to \pi^0 \pi^0 \pi^{\pm}$ 
decays are similar to the $K_L \to 2\pi$ decay
in that the EWP contribution cancels part
of the QCDP contribution. Due to this circumstance,
we suggest that a simultaneous study of the decays 
$K^{\pm} \to \pi^0 \pi^0 \pi^{\pm} $ and 
$ K^{\pm} \to  \pi^{\pm} \pi^{\pm}\pi^{\mp} $ 
 could throw new light on the relative strengths
 of the QCDP and
 the  EWP mechanisms in direct CP violation.

We shall estimate, in the framework  of 
the Standard Model,
the CP violating contributions to   the slope
parameters $g^+$ and $g^-$ characterizing the charged pion energy  
distributions in the $K^{\pm} \to \pi^0 \pi^0 \pi^{\pm}$ decays 
(formerly $\tau'$ decay).  The slope parameters are defined by 
the expansion
\begin{equation} 
|M(K^{\pm}(k) \to \pi^0(p_1) \pi^0 (p_2) \pi^{\pm}
(p_3))|^2 \propto 1+g^{\pm}Y+..., \label{slope-def}
\end{equation} 
where
\begin{equation}
Y=(s_3-s_0)/m^2_{\pi},\quad s_i=(k-p_i)^2, \quad s_0=m^2_K/3 +m^2_\pi.
\label{kinvar}
\end{equation}

Our tenet is corroborated by a calculation of
the amplitudes to leading non-vanishing order in a momentum expansion.  
As was previously found  for the $K^{\pm} \to \pi^{\pm} \pi^{\pm}
\pi^{\mp}$ decays \cite{2,3}, higher-order corrections do not  considerably
change the conclusion concerning the
relative magnitudes of the QCDP and EWP contributions to 
the difference $(g^+ -g^-)_{\tau}$. The role of higher-order corrections 
in the  ${\tau}'$ decays will be considered elsewhere.

\section{The $K^{\pm} \to \pi^0 \pi^0 \pi^{\pm} $ amplitude}

Our starting point is the 
 $\Delta S=1$ effective non-leptonic Lagrangian proposed in ref.~\cite{4}
\begin{equation}
L(\Delta S=1) =\sqrt{2}G_{\rm F}\sin\theta_{\rm C} \cos \theta_{\rm C}
\sum_i c_iO_i\,, \label{1} 
\end{equation} 
where $ O_{1-6}$ are  effective four-quark operators represented by the operator 
products 
\begin{align}
O_1=&\bar s_L\gamma_{\mu}d_L \cdot \bar u_L \gamma_{\mu}u_L-\bar s_L
\gamma_{\mu} u_L \cdot \bar u_L \gamma_{\mu}d_L \ ; \qquad(\{8_f\},
\Delta I=1/2), \\
O_2 =& \bar s_L \gamma_{\mu}d_L \cdot \bar u_L \gamma_{\mu} u_L +
\bar s_L \gamma_{\mu} u_L \cdot \bar u_L \gamma_{\mu} d_L +2 \bar s_L
\gamma_{\mu} d_L \cdot \bar d_L \gamma_{\mu} d_L \nonumber\\
  &+2 \bar s_L \gamma_{\mu} d_L \cdot \bar s_L \gamma_{\mu} s_L \ ; \qquad
(\{8_d\}, \Delta I=1/2),\\
O_3=&\bar s_L \gamma_{\mu} d_L \cdot \bar u_L \gamma_{\mu}u_L +\bar s_L
 \gamma_{\mu}u_L \cdot \bar u_L \gamma_{\mu} d_L +
                   2 \bar s_L \gamma_{\mu}
d_L \cdot \bar d_L \gamma_{\mu} d_L \nonumber\\
  & -3 \bar s_L \gamma_{\mu} d_L \cdot
\bar s_L \gamma_{\mu} s_L \ ; \qquad(\{27\}, \Delta I=1/2), \\
 O_4=&\bar s_L
\gamma_{\mu} d_L \cdot \bar u_L \gamma_{\mu} u_L + \bar s_L \gamma_{\mu}
u_L \cdot \bar u_L \gamma_{\mu} d_L \nonumber\\
  & -\bar s_L \gamma_{\mu} d_L \cdot
\bar d_L \gamma_{\mu} d_L\ ; \qquad (\{27\}, \Delta I =3/2),\\
 O_5=& \bar s_L \gamma_{\mu} \lambda^a
d_L(\sum_{q=u,d,s} \bar q_R \gamma_{\mu} \lambda^a q_R)\ ; \qquad (\{8\},
\Delta I=1/2),\\
O_6=&\bar s_L \gamma_{\mu} d_L(\sum_{q=u,d,s} \bar q_R \gamma_{\mu} q_R)
\ ; \qquad (\{8\}, \Delta I =1/2) \ .
\end{align}
Among these operators, only $O_4$ generates $\Delta I=3/2$ transitions.
The operators $O_{5,6}$ originate from the QCDP diagrams.
To calculate CP-violating effects, also the operators $ O_{7,8}$ 
generated by the EWP diagrams must be added,
\begin{align}
O_7=&\frac{3}{2} \bar s\gamma_{\mu}(1+\gamma_5)d (\sum_{q=u,d,s}e_q
\bar q \gamma_{\mu}(1-\gamma_5)q) ; \qquad  (\Delta I=1/2, 3/2),\\
O_8=&-12\sum_{q=u,d,s} e_q (\bar s_L
q_R)(\bar q_R d_L) ; \quad e_q=(\frac{2}{3}, -\frac{1}{3}, -\frac{1}{3});
\quad (\Delta I=1/2, 3/2). 
\end{align}
The corresponding Wilson coefficients $c_{7,8}$ are small, being
proportional to $\alpha_{em}$. The coefficients $c_{5-8}$ contain the
imaginary parts necessary for CP violation.

Bosonization of the operators $O_i$ is achieved through the substitutions
\cite{5}
\begin{align}
\bar q_j(1+\gamma_5)q_k =& -\frac{1}{\sqrt{2}}
F_{\pi}r(U-\frac{1}{\Lambda^2} \partial^2 U)_{kj}\ ,\\
\bar q_j \gamma_{\mu}(1+\gamma_5)q_k=& i[(\partial_{\mu} U)U^{\dag}
-U(\partial_{\mu}U^{\dag}) -\frac{rF_{\pi}}{\sqrt{2}
\Lambda^2}(m(\partial_{\mu}U^{\dag})-(\partial_{\mu}U)m)]_{kj}\ .
\end{align}
Here, $m$ is the diagonal quark-mass matrix,
$$
m={\rm Diag}\{m_u,m_d,m_s\} \ ,
$$
and the remaining parameters are defined as 
$$
r=2m^2_{\pi}/(m_u+m_d),\quad \Lambda \approx 1\; \mbox{GeV},\quad
F_{\pi}=93\; \mbox{MeV}.  $$ 
The $3\times3$ $U$-matrix is written as an expansion 
\begin{equation} 
U=\frac{F_{\pi}}{\sqrt{2}}\left(1+\frac{i\sqrt{2}\hat
\pi}{F_{\pi}}-\frac{\hat \pi^2}{F_{\pi}^2}+a_3\left(\frac{i\hat
\pi}{\sqrt{2} F_{\pi}}\right)^3 +2(a_3-1)\left(\frac{i \hat \pi}{\sqrt{2}
F_{\pi}}\right)^4 +....\right)\ 
\end{equation}
in the pseudoscalar nonet-meson-field matrix $\hat \pi$ 
\begin{equation}
\hat \pi=
\left (
\begin{array}{ccc}
\displaystyle{
\frac{\pi_0}{\sqrt{3}}+\frac{\pi_8}{\sqrt{6}}+\frac{\pi_3}{\sqrt{2}}  }&
\pi^+ & K^+ \\ \\
\pi^- &
\displaystyle{
\frac{\pi_0}{\sqrt{3}}+\frac{\pi_8}{\sqrt{6}}-\frac{\pi_3}{\sqrt{2}} } &
K^0 \\ \\
K^- & \bar K^0 & 
\displaystyle{  \frac{\pi_0}{\sqrt{3}}-\frac{2\pi_8}{\sqrt{6}}   }
\end{array}
\right)\ .
\end{equation}

The PCAC condition demands $a_3 =0$ \cite{6}  and we adopt this condition
 as well,
bearing in mind, that on mass shell, the values of the mesonic amplitudes
are independent of the parameter $a_3$.

In the calculation of the  $K\to 3 \pi$ amplitudes 
we make use of the Fierz identities for the colour matrices
\begin{align}
\delta^{\alpha}_{\beta}
\delta^{\gamma}_{\delta}=&\third \delta^{\alpha}_{\delta}
\delta^{\gamma}_{\beta} +\half \lambda^{\alpha}_{\delta}
\lambda^{\gamma}_{\beta}\nonumber \\ 
\lambda^{\alpha}_{\beta} \lambda^{\gamma}_{\delta}=&\extrabrak
\delta^{\alpha}_{\delta} \delta^{\gamma}_{\beta} -\third
\lambda^{\alpha}_{\delta} \lambda^{\gamma}_{\beta}\nonumber
\end{align}
as well as the  Fierz identities for the Dirac matrices
$$
\bar s \gamma_{\mu} (1+\gamma_5)d \cdot \bar q \gamma_{\mu} (1-\gamma_5)q=
-2\bar s (1-\gamma_5)q \cdot \bar q(1+\gamma_5)d \ .
$$
Thus, in  leading order non-vanishing approximation our result for the 
matrix element can be expressed as
\begin{align}
M\left(K^{\pm}\right.&\left. \to  \pi^0(p_1) \pi^0 (p_2) \pi^{\pm}(p_3)\right) 
    = \nonumber  \\
& = \kappa \left[1 \pm ia_{KM}
+\frac{3m^2_{\pi}}{m^2_K}\left(1-\frac{9c_4}{2c_0}\right) Y (1 \pm
ib_{KM})+...\right] \ .\label{K3pi} 
\end{align}
The kinematic variable $Y$ is defined in eq.(\ref{kinvar}). The overall 
strength is regulated by the parameter 
\begin{equation}
\kappa=\frac{G_Fm^2_K}{6\sqrt{2}} c_0\sin \theta_C \cos \theta_C \ ,
\end{equation}
and the remaining parameters are functions of the following combinations
\begin{equation}
c_0=c_1-c_2-c_3-c_4+\frac{32}{9} \beta \mbox{Re} \tilde c_5 \ ,
\end{equation}
\begin{equation} \tilde
c_5=c_5+\frac{3}{16}c_6, \quad \tilde c_7=c_7+3c_8 \ ,
\end{equation}
\begin{equation}
\beta=\frac{2m^4_{\pi}}{\Lambda^2 (m_u+m_d)^2} \ .
\end{equation}

The terms $a_{KM}$ and $b_{KM}$ in the amplitude 
(\ref{K3pi}) are the imaginary parts of the amplitude
generated by the Kobayashi-Maskawa phase. Explicitely, we find
\begin{eqnarray}
a_{KM}&=& \beta\left(\frac{32}{9}  \mbox{Im} \tilde c_5 +\frac{6
\Lambda^2 \mbox{Im}\tilde c_7}{m^2_K} \right)/c_0 \label{aKM-def}\\
b_{KM}&=& \beta \left( \frac{32}{9}  \mbox{Im} \tilde c_5 +\frac{3
\Lambda^2 \mbox{Im} \tilde c_7}{m^2_K-m^2_{\pi}}
\right)/(c_0-\frac{9c_4}{2}) \ , \label{bKM-def}
\end{eqnarray}
with coefficients as above.

Our approach can also be used to caculate the $K\to 2\pi$ amplitudes.
For their real parts we get
\begin{eqnarray}
M(K^0_1 \to \pi^+ \pi^-)& =&\frac{G_F F_{\pi}}{\sqrt{2}}\sin \theta_C \cos
\theta_C (m^2_K-m^2_{\pi}) c_0 \ , \label{K2pi1} \\
M(K^+ \to \pi^+ \pi^0)&=&\frac{G_F F_{\pi}}{\sqrt{2}} \sin \theta_C \cos
\theta_C (m^2_K-m^2_{\pi}) (\threehalf c_4 ) \ .
\end{eqnarray}
A comparison between the real parts of the amplitudes 
of eqs (\ref{K3pi}) and (\ref{K2pi1}) shows that their ratio
is nothing more than a reflection of the well-known relation
\begin{equation}
\begin{array}{rrr}
M\left(K^+(k)\to \pi^0(p_1) \pi^0(p_2) \pi^+(p_3)\right)
=\displaystyle{ \frac{1}{6F_{\pi}}  } M(K^0_1 \to
\pi^+\pi^-)\\ \\
\times \left[1+  \displaystyle{ 
\frac{3m^2_{\pi}}{m^2_K} \left(1-\frac{3M(K^+ \to
\pi^+\pi^0)}{M(K^0_1 \to \pi^+\pi^-)}   \right) }   Y  
 \right]
\end{array}
\end{equation}
obtained earlier 
\cite{7,8,9}\footnote{In \cite{9} there is a
misprint: a factor $y$ is missing after the factor $(1+
\frac{3\delta}{1+ \theta})$ in the expression for  the $K^+ \to \pi^0
\pi^0 \pi^+$ amplitude, eq.(6.9).} using soft-pion techniques and current 
algebra.

From the data on the  $K\to 2\pi$ decay rates  \cite{10}, 
it follows that 
\begin{eqnarray}
c_1-c_2-c_3 +\frac{32}{9}\beta\, \mbox{Re} \tilde c_5& =& -10.13 \ ,\\
c_4&=&0.328 \     .
\end{eqnarray}
Furthermore, the combination $\beta\, \mbox{Re} \tilde c_5$ 
can be determined separately, provided we are willing to accept the
 estimate of Shifman {\it et al.}~\cite{4,11},
\begin{equation}
c_1-c_2-c_3 =-2.89 \ ,
\end{equation}
which leads to the value
\begin{equation}
\frac{32}{9}\beta\, \mbox{Re} \tilde c_5 = -7.24\ .
\end{equation}

To estimate CP-odd effects in $K^{\pm} \to (3\pi)^{\pm}$ decays, 
described 
by the coefficients $a_{KM}$ and by $b_{KM}$,
we need a certain combination of the parameters $\mbox{Im} \tilde c_5$ 
and $\mbox{Im} \tilde c_7$.  The theoretical preditions for these
parameters are very uncertain and different authors
(see \cite{3}) give different results. Fortunately, the combination 
entering the  $K^{\pm} \to \pi^0 \pi^0 \pi^{\pm}$
amplitude turns out to be similar to the combination determining the
parameter $\varepsilon'$ in $K_L \to 2\pi$ decay \cite{2}. 
This circumstance allows us to 
obtain a reliable estimate of $(g^+ -g^-)_{\tau'}$.

\section{Estimate of the CP-odd difference $(g^+ -g^-)_{\tau'}$}

Although the amplitude (\ref{K3pi})  incorporates the imaginary 
terms necessary for
CP violation, this is  not sufficient for producing observable 
CP-violating  effects. In fact, the observable effects
arise from the interference between  these CP-odd terms and the 
CP-even imaginary terms created by the strong-interaction final-state 
rescattering between the pions. The strong-interaction effects are 
introduced into the
$K\to 3\pi$ amplitudes of eq.(\ref{K3pi}) by adding two terms,
 $a_{\tau'}$ and $b_{\tau'}$, so that
\begin{equation}
\begin{array}{rrr}
M\left(K^{\pm} (k) \to \pi^0(p_1) \pi^0(p_2) \pi^{\pm}(p_3) \right)=\kappa
\displaystyle{ \frac{1\pm ia_{KM}}{(1+a^2_{KM})^{1/2}}  }
 [1+ia_{\tau'} + \\ \\
+\displaystyle{ \frac{3m^2_{\pi}}{m^2_K} \left(1-\frac{9c_4}{2c_0}  
 \right)  }    Y(1+ib_{\tau'} \pm
i(b_{KM}-a_{KM}))+...]\ .\label{Kto3pi-full}
\end{array}
\end{equation}
This assumption is valid as long as the rescattering contribution
can be treated in the linear approximation.

The slope parameters $g^+$ and $g^-$ were defined in 
eq.(\ref{slope-def}).
From this definition and  eq.(\ref{Kto3pi-full}) we get for the 
relative difference in slope parameter for the $K\to 3\pi$ decays
\begin{equation}
\Delta g_{\tau'}=\left(\frac{g^+ -g^-}{g^++ g^-} \right)_{\tau'}=
\frac{a_{\tau'}(b_{KM}-a_{KM})_{\tau'}}{1+a_{\tau'}b_{\tau'}} \label{Deltag}
\end{equation}
The strong-interaction-rescattering parameters, 
$a_{\tau'}$ and $b_{\tau'}$, are
determined by calculating the imaginary parts of the loop diagrams of 
Fig.~1. Putting the intermediate pions on shell (see Appendix)
yields, in leading approximation, 
\begin{equation}
a_{\tau'}=0.12, \qquad b_{\tau'}=0.49 .  
\end{equation} 
The CP-odd numerator of eq.(\ref{Deltag})  
can be calculated from the expressions
in eqs (\ref{aKM-def}) and  (\ref{bKM-def}), and is found 
being equal to
\begin{eqnarray}
(b_{KM}-a_{KM})_{\tau'}&=&\frac{16c_4\beta\;
\mbox{Im} \tilde c_5 }{c_0(c_0-\ninehalf c_4)}
 -\frac{6\beta \Lambda^2\; \mbox{Im} \tilde c_7}{m^2_K
c_0} \left(1-\frac{c_0m^2_K}{2(m^2_K-m^2_{\pi})(c_0-\ninehalf c_4)}
\right) \nonumber \\ &&  \nonumber \\
 &=& 0.042 \beta\, \mbox{Im} \tilde c_5\, (1+27.8\; \mbox{Im} 
\tilde c_7/\mbox{Im} \tilde c_5) \ .\label{abCP-odd}
\end{eqnarray}
The combination of Wilson coefficients in this formula,
\begin{equation}
\beta\; \mbox{Im}\tilde c_5\,(1+27.8\; \mbox{Im} \tilde c_7/\mbox{Im} \tilde
c_5) \ ,\label{Comb3pi}
\end{equation}
is very similar to  another combination
\begin{equation}
\beta\, \mbox{Im} \tilde c_5(1+\frac{24.36}{1-\Omega} \cdot
\frac{\mbox{Im}\tilde c_7}{\mbox{Im} \tilde c_5}) =-\frac{(1.63 \pm
   0.25)\cdot10^{-4}}{1-\Omega} \beta\, \mbox{Re} \tilde c_5
\label{Comb2pi}
\end{equation}
defining the direct CP-violating parameter $\varepsilon'$ in  
$K_L \to 2\pi$ decay \cite{2,3}. 
The parameter $\Omega$ takes into account  isospin-breaking 
contributions generated by the two-step transition  $K^0
\to \pi^0 \eta(\eta') \to \pi^0 \pi^0$.

At $\Omega=0.124$  expressions (\ref{Comb3pi}) and (\ref{Comb2pi}) 
coincide, giving
\begin{equation}
\Delta g_{\tau'}=(1.8 \pm 0.28)\cdot 10^{-6} \ ,
\end{equation}
and at $\Omega$=0.25
\begin{equation}
\Delta g_{\tau'}=2.1\cdot 10^{-6}(1-\frac{4.7\, \mbox{Im} \tilde
c_7/\mbox{Im} \tilde c_5}{1+32.48\, \mbox{Im} \tilde c_7/\mbox{Im}
\tilde c_5})\ .  
\end{equation} 
Both values of
$\Omega$ are in line with estimates figurating in the literature
(see \cite{12} and references therein).

\section{The CP-odd difference $(g^+ -g^-)_{\tau}$}

As we shall now show, our result for the slope-parameter 
difference  in $\tau'$ decay, as embodied in 
 eq.(\ref{abCP-odd}), enables us to draw quite precise conclusions 
concerning the magnitude of
the slope-parameter difference in another decay, namely the
$K^{\pm} \to \pi^{\pm}\pi^{\pm}\pi^{\mp}$ decay, or  
 $\tau$ decay. From eqs (33) and (34) in ref.~\cite{3}, 
we derive the following relation  
\begin{equation}
(b_{KM} -a_{KM})_{\tau}=-2\left[ \frac{16c_4\beta\,
\mbox{Im} \tilde c_5 }{c_0(c_0+9c_4)}
  + \frac{3\beta \Lambda^2\, \mbox{Im}
{\tilde c}_7 }{m^2_K c_0}
\left(1+\frac{12c_4 m^2_K}{\Lambda^2 (c_0+9c_4)} \right) \right] \ .
\label{abCP-odd-tau} 
\end{equation}
The slope-parameter difference $\Delta g_{\tau}$ is again given by
expression  (\ref{Deltag}), provided index $\tau'$ is everywhere
replaced by   $\tau$. The value of the rescattering parameter
$a$ does not change, but that of $b$ does,
\begin{equation}
a_{\tau}=0.12, \qquad b_{\tau}=0.714 .  
\end{equation} 

Combining eqs  (\ref{Deltag}), (\ref{abCP-odd}) and  (\ref{abCP-odd-tau})
we can form the ratio of the slope parameters differences, 
\begin{equation} 
\frac{-\Delta g_{\tau}}{\Delta
g_{\tau'}}=2\frac{c_0-9c_4/2}{c_0+9c_4} \cdot \frac{1-14.34\,
\mbox{Im} \tilde c_7/\mbox{Im} \tilde c_5}{1+ 27.8\, \mbox{Im}\tilde
c_7/\mbox{Im} \tilde c_5}\cdot 
  \frac{1+a_{\tau'}b_{\tau'}}{1+a_{\tau}b_{\tau}}\ .\label{Dgprim/Dg}
\end{equation} 
Now, only negative values of the ratio $\mbox{Im} \tilde
c_7/\mbox{Im} \tilde c_5$ appear in the literature \cite{3}.
If furthermore, we assume that the numerical value of
this ratio is so small that the sign of the right hand side
of eq.(\ref{Dgprim/Dg}) is positive,
we may conclude that
\begin{equation} -\Delta g_{\tau} \ge 3.1 \Delta g_{\tau'} > 0.56
\cdot 10^{-5}.
\end{equation} 
This result differs from other
estimates,  as exemplified by refs.~\cite{13} 
and  \cite{14}
\begin{equation} -\Delta g_{\tau}=1.8 \Delta
g_{\tau'} \quad \cite{13},\qquad - \Delta g _{\tau}=2.2 \Delta
g_{\tau'} \quad \cite{14}.  
\end{equation} 

Moreover, as follows from our discussion in Sect.~3, 
we strongly believe $\Delta g_{\tau'}$ to be 
of order $ 10^{-6}$. In contrast,  $\Delta g_{\tau}$ can
reach  values of order  $10^{-5}$ , providing the EWP
contribution cancels out a considerable part of the
 QCDP contribution (see eq.(\ref{Dgprim/Dg})).  
For example, if the EWP cancels half of the
QCDP contribution, then 
\begin{equation} -\Delta g_{\tau} =7.8
\Delta g_{\tau'} \ge 1.4 \cdot 10^{-5} 
\end{equation} 
and if the EWP
cancels three-quarters of the QCDP contribution, then 
\begin{equation}
-\Delta g_{\tau}=17.2 \Delta g_{\tau'}\ge 3.1 \cdot 10^{-5} .
\end{equation}

\begin{center}
\begin{table}[h]
\begin{tabular}{|c|c|c|}
\hline \quad $\Delta g_{\tau}$ (in units $10^{-5}$) & $\Delta g_{\tau'}$
(in units $10^{-5}$) & Refs. \\ \hline \hline $-700 \pm 500$ &  $-15 \pm
275$ & [15] \\ \hline $|\Delta g_{\tau}|_{LO} \le 0.7$ &  - & [16]
\\ \hline -0.16 & - & [17] \\ \hline $|\Delta g_{\tau}|=38.2$ &
$|\Delta g_{\tau'}|=31.5$ &  [18]    \\ \hline $-0.23 \pm 0.06$ &
$0.13 \pm 0.04$ & [13] \\ \hline $(-4.9 \pm 0.9)\sin \delta$ & - &
[2] \\ \hline $-2.4 \pm 1.2$ & $1.1 \pm 0.7$ & [14]
\\ \hline $-(3.0 \pm 0.5)x; \quad  0.5< x < 5.0$ & - & [3] \\ \hline
 \hline
$(- \Delta g_{\tau})_{LO}>(0.56\pm0.09) f(x)$   & $0.18 \pm 0.03$    &
present
\\ At $x=1, \quad  (-\Delta g_{\tau}) =2.9 \pm 0.6$ &      &  work
\\ \hline
\end{tabular}
\caption{ Values for the slope-parameter ratios
$\Delta g_{\tau}$ and  $\Delta g_{\tau'}$ in $\tau$ and $\tau'$ decays,
 in units of $10^{-5}$.}
\end{table}
\end{center}

These examples show that a simultaneous measurement of $\Delta
g_{\tau'}$ and $\Delta g_{\tau}$ can clear up the question 
about the true relative
strength of EWP and QCDP mechanisms in direct CP violation. The
estimates of the values $\Delta g_{\tau}$ and $\Delta g_{\tau'}$
as obtained in other investigations  are summarised in  Table 1.

\section{Concluding remarks}

We have calculated the CP-odd difference of  slope parameters,
 $\Delta g_{\tau'}$ of  eq.(\ref{Deltag}),  in the  ${\tau'}$ decays 
 $ K^{\pm} \to  \pi^0 \pi^0 \pi^{\pm}$
in leading non-vanishing approximation in a momentum expansion
of the decay amplitude. 

We observe that the difference of  slope parameters $\Delta
g_{\tau'}$ in  $ K^{\pm} \to  \pi^0 \pi^0 \pi^{\pm}$ decay 
and the parameter $\varepsilon'$ in $K_L \to2\pi$ decay both depend
practically on one and the same combination of the Wilson 
coefficients $\mbox{Im} \tilde c_5$ and
$\mbox{Im} \tilde c_7$. This observation permits a reliable estimate 
of  $\Delta g_{\tau'}$
using the known magnitude of $\varepsilon'$.

A comparison with the value of the
corresponding parameter $\Delta g_{\tau}$  in the  ${\tau}$ decays 
 $K^{\pm} \to \pi^{\pm} \pi^{\pm} \pi^{\mp}$ shows that
 $\Delta g_{\tau}$ is expected to be at least 3 times larger than
 $\Delta g_{\tau}'$. 
In fact, it  may be even one order of magnitude 
larger, provided there is a sizeable cancellation between 
the electroweak-penguin  and the QCD-penguin   contributions
to the parameter $\varepsilon'$. Such a cancellation
is not excluded  \cite{3,19}.

 We have not considered the possibility of a sequential decay
 $K^{\pm}\to \pi^0 \eta \pi^{\pm} \to \pi^0 \pi^0 \pi^{\pm}$
through an intermediate $ \eta \to \pi^0$ transition, a correction
which is of order $p^4$. 
We shall study this possibility elsewhere. 
In the case of $K^{\pm} \to \pi^{\pm} \pi^{\pm}
\pi^{\mp}$ decay, higher-order corrections increase  $\Delta
g_{\tau}$ by 20\%, but change  very little the relation between 
 electroweak-penguin and QCD-penguin 
contributions \cite{3}. We expect a similar increase of 
$\Delta g_{\tau'}$ in the 
$K^{\pm} \to \pi^0 \pi^0 \pi^{\pm} $ decay, 
since $a_{\tau'}\approx a_{\tau}$, and according to  ref.\cite{3}
 $p^4$ corrections increase the value of $a_{\tau}$ by 30\%.

\vspace{1cm}
{\large \bf Acknowledgments.}
We would like to thank the Swedish Research Council for financial support.
One of us (E.Sh) would also like to acknowledge a partial financial support
from the Grant RFBR-02-02 16957.
 
\section{Appendix}

Here, we shall calculate the CP-even imaginary part coming from the
pion-rescattering diagrams displayed in fig.~1. The imaginary
part of a diagram is obtained by cutting the internal lines as shown.

\begin{figure}[h]
  \begin{tabular}{c@{}c@{}c@{}}
\scalebox{.70}{\includegraphics{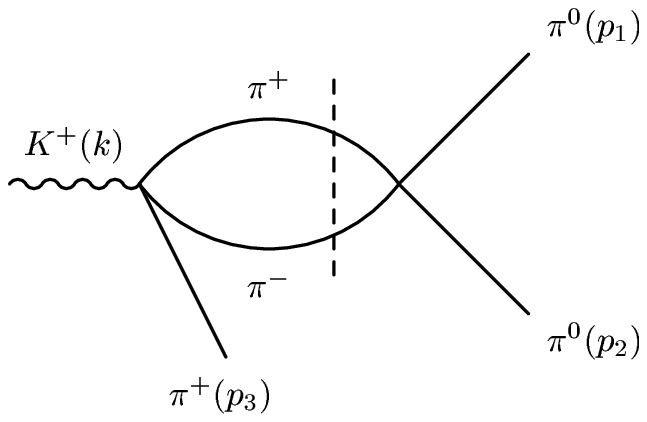}}&
\scalebox{.70}{\includegraphics{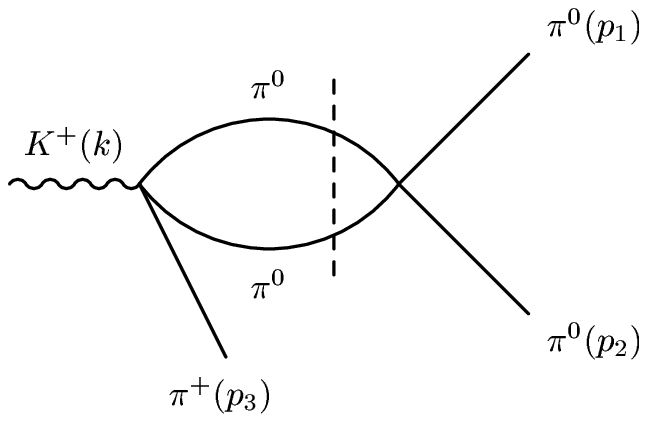}}&
\scalebox{.70}{\includegraphics{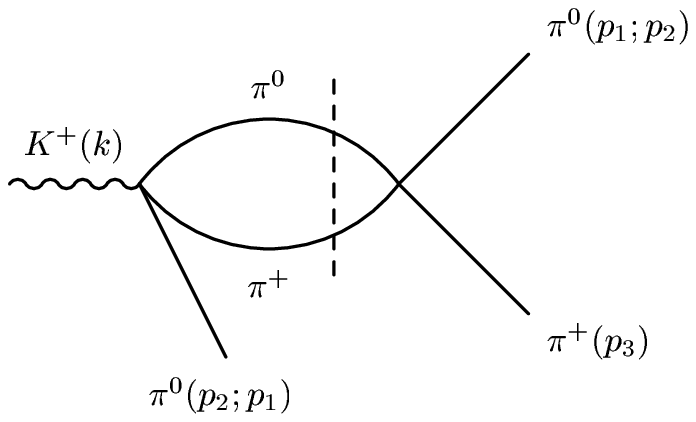}}  \\
a) &
b) &
c)+d)
  \end{tabular}
\caption{Rescattering diagrams for the imaginary part.
Diagrams are cut along the dashed line. Diagrams c) and d)
are related through $ \pi^0(p_1)\leftrightarrow
\pi^0(p_2)$.}
\end{figure}

We start from the amplitudes
\begin{eqnarray}
M\left(K^+(k) \to  \pi^0(p_1) \pi^0 (p_2) \pi^+(p_3)\right)
     &=&A+B(s_0-s_3) \label{tau-prime}\\
M\left(K^+(k) \to  \pi^+(p_1) \pi^+ (p_2) \pi^-(p_3)\right)
 &=&A'+B'(s_0-s_3) \label{tau-no-prime}
\end{eqnarray}
with kinematic variables as defined in eq.(\ref{kinvar}). The $\tau'$
decay amplitude, eq.(\ref{tau-prime}), is given in eq.(\ref{K3pi}), 
of which we only need the leading real term. 
In  the $\tau$ decay amplitude, eq.(\ref{tau-no-prime}), 
the parameter $A'$ is twice as large as $A$. 
 For the $\pi \pi$ scattering amplitudes we
insert the leading-order approximations,
\begin{eqnarray}
M\left( \pi^+(q_1)\pi^-(q_2) \to  \pi^0(q_3)\pi^0(q_4)\right)
     &=&\frac{i}{F_{\pi}^2}(s-m_{\pi}^2) \\
M\left( \pi^0(q_1)\pi^0(q_2) \to  \pi^0(q_3)\pi^0(q_4)\right)
     &=&\frac{i}{F_{\pi}^2}(s+t+u-3m_{\pi}^2) \\
M\left( \pi^0(q_1)\pi^+(q_2) \to  \pi^0(q_3)\pi^+(q_4)\right)
     &=&\frac{i}{F_{\pi}^2}(t-m_{\pi}^2)
\end{eqnarray}
with $s=(q_1+q_2)^2,$  $t=(q_1-q_3)^2$ and  $u=(q_1-q_4)^2$ 
as usual.

First, we calculate diagram a) with a $\pi^+\pi^-$ pair in
the loop. The result is an imaginary  contribution to the
$\tau'$ decay
\begin{equation}
\delta M_a=\frac{i}{16\pi F_{\pi}^2}(s_3-\mu^2)
 \sqrt{1-\frac{4\mu^2}{s_3}}
\left[ A'+\half B'(s_3-s_0) \right] \ . \label{diag-a-exact}
\end{equation}
However, we are not interested in the exact value of 
$\delta M_a$. The slope parameters, eq.(\ref{slope-def}),
 are defined through an expansion in $Y=(s_3-s_0)/m_{\pi}^2$.
Moreover, we normalise the $K^+\to\pi^+\pi^+\pi^-$ decay 
parameters as
\begin{eqnarray}
A'&=&\twothird m_K^2 \\
B'&=&1+9c_4/c_0=0.718\ ,
\end{eqnarray}
so that a short algebraic calculation gives as result
\begin{equation}
\delta M_a=i\frac{m_K^4}{72\pi F_{\pi}^2} 
   \sqrt{\frac{s_0-4\mu^2}{s_0}} 
  \left[ 1+ \frac{3(s_3-s_0)}{m_K^2} 
 \left\{1+\frac{2m_K^2m_{\pi}^2}{3s_0(s_0-4m_{\pi}^2)}
   + \fourth B'\right\} \right]  \ .
\end{equation}

Diagram b) with two neutral pions in the  loop give a contribution to
the imaginary part
\begin{equation}
\delta M_b=\frac{i}{32\pi F_{\pi}^2}\mu^2
 \sqrt{1-\frac{4\mu^2}{s_3}}
\left[ A+ B(s_0-s_3) \right] \ .\label{diag-b-exact} 
\end{equation}
The parameters for the decay  $K^+\to\pi^+\pi^0\pi^0$  are
\begin{eqnarray}
A&=&\third m_K^2 \\
B&=&-(1-9c_4/2c_0)= -1.14\ .
\end{eqnarray}
The expansion of this contribution yields the result
\begin{equation}
\delta M_b=i\frac{m_K^2 m_{\pi}^2}{96\pi F_{\pi}^2} 
   \sqrt{\frac{s_0-4\mu^2}{s_0}} 
  \left[ 1+ \frac{s_3-s_0}{m_K^2} 
  \left\{\frac{2m_K^2m_{\pi}^2}{3s_0(s_0-4m_{\pi}^2)}
   - 3 B\right\} \right]  \ .
\end{equation}

There are two contributions, diagrams c) and d), 
with $\pi^+\pi^0$ in the loop, since the final
state is symmetric in the two neutral pions, $\pi^0(p_1)$ and 
 $\pi^0(p_2)$. We shall not give the exact expressions,
corresponding to eqs (\ref{diag-a-exact}) and 
(\ref{diag-b-exact}), since they are
somewhat complicated. The expansion of the sum of the two 
amplitudes results in an imaginary contribution 
\begin{align}
\delta M_c+\delta M_d  = &
i\frac{m_K^4}{144\pi F_{\pi}^2} 
   \sqrt{\frac{s_0-4\mu^2}{s_0}} 
  \left[ -(1-\frac{3m_{\pi}^2}{m_K^2}) \right. \nonumber \\ 
 & + \left.\frac{3(s_3-s_0)}{2m_K^2} 
  \left\{1+\frac{2m_{\pi}^2(s_0-2m_{\pi}^2)}{s_0(s_0-4m_{\pi}^2)}
   + \frac{3m_{\pi}^2}{m_K^2} B\right\} \right]  \ .
\end{align}


\begin{thebibliography}{99}
\bibitem{1} Flynn, J.M.~and Randall, L., Phys.Lett.~{\bf B224}, 221 (1989);
Buchalla, G.~{\it et al}., Nucl.Phys.~{\bf B337}, 313 (1990); 
 Paschos, E.A.~and Wu, Y.L., Mod.Phys.Lett.~{\bf A6}, 93 (1991).
\bibitem{2} Shabalin, E., Proc. of {\it Les Rencontres de Physique de la
Vallee d'Aoste}, La Thuile, Aosta Valley, March 9-15, 2003, p.417;
hep-ph/0305320.
\bibitem{3} Shabalin, E., Report at {\it Physics at Meson
Factories, DAPhNE'2004}, Frascati, June 7-11, 2004; hep-ph/0405229; 
  Phys.Atomic Nucl.~{\bf 68}, 88 (2005).
\bibitem{4} Shifman, M.A., Vainshtein, A.I.~and Zakharov, V.I., 
 Zh.Eksp.Teor.Fiz.~{\bf 72}, 1275 (1977); 
   Sov.Phys.JETP~{\bf 45}, 670 (1977).
\bibitem{5} Bardeen, W.A., Buras, A.J.~and G{\'e}rard, J.-M., 
   Nucl.Phys.~{\bf B293}, 787 (1987).
\bibitem{6} Cronin, J., Phys.Rev.~{\bf 161}, 1483 (1967).
\bibitem{7} Bouchiat, C.~and Meyer, Ph., Phys.Lett.~{\bf B25}, 282 (1967).
\bibitem{8} Dolgov, A.D.~and Zakharov, V.I., Yad.Fiz.~{\bf 7}, 352 (1968);
   JETP Lett.~{\bf 7}, 323  (1968).
\bibitem{9} Vainshtein, A.I.~and Zakharov, V.I., 
         Sov.Phys.Uspekhi, {\bf 13}, 73 (1970).
\bibitem{10} Shabalin, E., Nucl.Phys.~{\bf B409}, 87 (1993).
\bibitem{11} Okun, L.B., {\it Leptons and Quarks}  (North-Holland, 1982), 
  p. 57.
\bibitem{12} Cirigliano, V.~{\it et al}., Eur.Phys.J.~{\bf C33}, 369 (2004);
Bertolini S., Fabbrichesi M.~and Eeg J.O., Rev.Mod.Phys.~{\bf 72}, 65 (2000).
\bibitem{13} Maiani, L.~and Paver, N., In {\it The Second DAPhNE Physics
Handbook}, Eds L.Maiani, G.Pancheri and N.Paver (INFN-LNF, 1995), p.51.
\bibitem{14} Scimemi, I., Gamiz, E.~and Prades, J., hep-ph/0405204.
\bibitem{15} Bel'kov, A.A.~{\it et al}., Phys.Lett.~{\bf B232}, 118 (1989).
\bibitem{16} D'Ambrosio, G., Isidori, G.~and Paver, N., Phys.Lett.~{\bf B273},
         497 (1991).
\bibitem{17} Isidori, G., Maiani, L.~and Pugliese, A., Nucl.Phys.~{\bf B381}, 
                 522 (1992).
\bibitem{18} Bel'kov, A.A.~{\it  et al}., Phys.Lett.~{\bf B300}, 283 (1993).
\bibitem{19} Hambye, T., Peris, S.~and de Rafael, E., 
    JHEP {\bf 0305}, 027 (2003);
Bertolini, S., Eeg, J.O.~and Fabbrichese, M., Phys.Rev.~{\bf D63}, 056009
(2001); Donoghue, J.F.~and Golowich, E., Phys.Lett.~{\bf B478}, 172
(2000).

\end{thebibliography}
\end{document}